# 'No, auntie, that's false': Female baby boomers develop critical skills to confront fake news with guidance from relatives

Andrea Pecho-Ninapaytan, Mtr
Stefany Zambrano-Zuta, Mtr
Lizardo Vargas-Bianchi, PhD
Universidad de Lima, Graduate School

**Abstract**
The spread of fake news has been increasing, which gives rise to a special interest in the development of identification and coping skills among news consumers so that they can filter out misleading information. Studies suggest that older people share more fake news from social media. There is scarce literature that analyse how baby boomers behave in the face of fake news. The purpose of this study is to examine how female baby boomers deal with fake news on Facebook and their available resources to learn how to identify and handle dubious information. A qualitative study and thematic analysis were conducted using information obtained from interviewing female baby boomers. Four themes emerge from the analysis, revealing that participants recognise that they can identify fake news but may not always be able to do so due to limitations in their understanding of an issue or uncertainty about its source. Participants show participants empirically develop critical identification and filtering skills with the assistance from close family members.

**Introduction**
The spread of fake news has become a global concern for citizens, media and governments (Talwar, 2020), as people adopt misperceptions and misconceptions and spread them to others (Buchanan & Benson, 2019). Even news media are susceptible to sharing false information (Boyd, 2017). Authors argue that social media dissemination of fake news has contributed to poor health decisions (Hotez, 2016), changed the direction of the stock market (Ferrara, et al., 2016), and even affected the credibility of brands of products and services (Visentin, Pizzi & Pichierri, 2019). The reasons people share fake news are not precisely known (Talwar, Dhir, Kaur, Zafar & Alrasheedy, 2019). Studies attribute it to low critical interpretation of information, difficulty in following the news flow and lack of skill with internet tools (Yabrude, Souza, Campos, Bohn & Tiboni, 2020).

Technology does not stop developing and presents communication tools, which we gradually learn to use and adopt. We usually associate technology adoption with youth, besides their ability to understand the structure of devices and applications better than older people. Niemelä (2007) argues that baby boomers have grown up along with advances in technology, and that this reality makes it easier for older adults not always to perceive digital tools as something distant. However, an analysis of the spread of fake news during the 2016 presidential election in North America identified that the Facebook users most likely to share fake news were adults over the age of 65. People in that age group, the baby boomers, shared seven times more fake news articles on Facebook than younger users

(Guess, Nagler & Tucker 2019). Guess, Nyhan, & Reifler (2018) note that Facebook contains the highest volume of fake news presence. The literature on how baby boomers conduct themselves in the face of fake news, in particular their ability to identify fake news, is limited (Loos & Nijenhuis, 2020). This study aims to analyse how women belonging to this age group deal with fake news found on the social network Facebook, and how they deal with it.

During the initial stage of the study, the aim of the study was to identify the filters that baby bombers possess to identify misleading information on the social network. However, during the preliminary analysis of the data, it was revealed that individuals not only possessed filters, but that they developed them based on the learning and experiences they acquired from close family members. Based on this finding, the questions were reformulated:

> RQ1: What is the experience of baby boomer women when they are confronted with and have to deal with the fake news they find on the social network Facebook?
>
> RQ2: How do the people close to baby boomer women contribute to their identification and coping skills when they face the fake news they find on the social networking site Facebook?

The research contributes to the advancement of knowledge about the process of media literacy that allows baby boomer women, who do not consciously seek to disseminate misleading information, to cope with fake news on social networks. Examining this phenomenon is of interest because access to accurate information improves people's quality of life (Loos & Nijenhuis, 2020), particularly during events of political uncertainty or health emergencies such as the Covid-19 pandemic.

**Background literature**
Using social media to deliver fake news has catalysed its dissemination, to the extent that more fake news articles are shared on social media than real news articles (Silverman, 2016). This phenomenon is observed among Facebook users, who are more likely to find and share this type of content across their contacts (Guess, et al., 2018). The design of this social platform prioritises popular posts, which increases their visibility and makes it difficult to differentiate between serious and misleading content (Salganik, et al., 2006; Hodas & Lerman, 2012; Nematzadeh, et al., 2017). Pariser (2011) argues that fake news is just as likely to go viral as genuine news. However, the sheer volume of false news makes it difficult to separate truth from fiction.

Some studies point out that fake news has particular characteristics. For example, they include first-person and second-person pronouns (Rashkin, et al., 2017; Ott and Rayson, 2011). In addition, they show a use of superlatives to exaggerate misleading content, a low prevalence of comparative and quantitative data (Ott and Rayson, 2011). Authors have observed that news consumption on networks can be reduced to specific topics and the formation of groups of people around those topics. This consumption practice confirms individuals' views, predisposing them to polarisation and to dismiss messages that contradict their perspectives (Moravec, Minas & Dennis, 2018; Guess, 2015; Guess, et al., 2018; Green, et al., 2002; Klofstad, 2009; Schäffer, 2007).

Regarding people's ability to deal with fake news, Buchanan & Benson (2019) concluded that besides the properties of a news source, receiver characteristics affect the likelihood of people spreading false information in social networks. Lutzke, Drummond, Slovic & Árvai (2019) analysed the dissemination of false information on Facebook about climate issues. The researchers identified that exercising critical thinking reduces the likelihood of trusting, liking and sharing fake news, without affecting individuals' recognition of reliable information. Another study is consistent with the latter findings, which reveals that people use a variety of strategies to filter out false information, such as personal judgements about the plausibility of a news story, and scepticism about their sources and journalistic style Flintham et al. (2018).

Wardle (2017) argues that there are different factors that favour the elaboration of fake news, and its underlying motivations to disseminate them. For example, poor journalism practice, parody to provoke, partisan posturing, seeking to exploit or gain power, or political influence and propaganda. Osmundsen, Bor, Vahlstrup, Bechmann & Petersen (2020) made similar findings in their study, claiming that the sharing of fake news reflects an inability to discern whether information is true or false. Their main conclusion was that the exchange of fake news, like the exchange of actual news, reflects partisan objectives. Resources such as artificial intelligence are also available to produce information of this nature. Bots programmed as a user on a social network act by algorithms to generate content and interact with humans, or other bot users, increasing the incidence of misleading news (Ferrara, Varol, Davis, Menczer, & Flammini, 2016; Chu, Gianvecchio, Wang, & Jajodia, 2012). These technologies are also used for stopping fake news (Della Vedova et al., 2018).

The post-war generation known as baby boomers grew up with the development of technological advances in communication, such as television or the telephone, and have shown evidence of incorporating the learning of new digital technologies (Venter, 2017). For example, they are no strangers to email, searching for information on the web, the use of social networks and other computer-mediated communication media. Prensky (2001) has called this generation 'digital immigrants' since digital is their second language, in contrast to more recent generations. Adolfsson, Strömberg & Stenberg (2017) found that generational affiliation does not affect public trust in media. However, baby boomers are less confident in their ability to identify the falsity of a news story, when compared to young millennials. Regardless of the generation to which one belongs, fake news is a phenomenon that affects everyone in different dimensions of life (Hotez, 2016; Ferrara, et al., 2016; Visentin, Pizzi & Pichierri, 2019).

There has been insufficient research on the role of age in the consumption of fake news on social media (Loos & Nijenhuis, 2020). Research that has analysed the dissemination of fake news and media literacy has not included age as one of the mediating factors. Available studies focus on younger populations who are already familiar with technology. O examines the psychological variables underlying the consumption and dissemination of misleading information, regardless of the age of those involved (Pennycook, Cannon & Rand, 2018). However, non-digital natives, and older adults, are a vulnerable population for online risks. It is suggested that they are populations whose limited media literacy may explain this vulnerability (Lee, 2018).

Some studies have identified that older adults may be more likely to share fake news (Guess, Nagler & Tucker 2019). Limited analytical reasoning skills may explain the challenges faced by adults in the face of fake news. Pehlivanoglu et al (2021) identified that although dogmatic individuals and religious fundamentalists were more likely to believe fake news, these relationships were partially or wholly explained by lower levels of analytical reasoning. The authors show that greater analytical reasoning was associated with greater accuracy in identifying fake news, while analytical reasoning was not associated with greater accuracy in detecting genuine news. However, this is contradicted by other research. It could be argued that these individuals have more developed knowledge networks, which increases the likelihood that they will apply the relevant knowledge at the moment (Newman & Zhang, 2021). The authors argue that older adults have more developed knowledge networks, which increases the likelihood that they can apply relevant knowledge when faced with misleading information. Another variable that may moderate the ability to identify fake news is the degree to which adults are exposed to misleading information. Facing little fake news would not seem to be sufficient to form critical skills to identify it. However, authors have found that even a single exposure to fake news on Facebook increases subsequent perceptions of accuracy among individuals (Pennycook, Cannon & Rand, 2018).

**Methods**
The study followed an inductive method. A qualitative method with a phenomenological design was used to address the research objectives. Phenomenological design incorporates individuals' worldviews, rational understandings and subjective meanings, and allows for understanding, exploring and describing what the individuals studied have in common regarding their experiences with a phenomenon (Creswell, 2013b; Wertz et al., 2011; Norlyk & Harder, 2010; Esbensen, Swane, Hallberg, & Thome, 2008; Kvåle, 2007; Graneheim & Lundman, 2004). The phenomenon under study can manifest itself as feelings, emotions, visions, perceptions, reasoning, etc. (Benner, 2008; Alvarez-Gayou, 2003; Bogden & Biklen, 2003; and Patton, 2002).

**Sample**. The sample was composed using a snowball or referral technique (Daymon and Holloway, 2010). This technique was appropriate for identifying participants within the context of the constraints posed by the context of the COVID-19 pandemic and the quarantine that followed. The study participants were women, aged 65-75 years, living in an urban setting in a western capital city. Palinkas et al. (2015) argue that qualitative sampling strategies should aim to select individuals, or data sources, considered 'information rich'. For this purpose, the following inclusion criteria were established: (a) they have a personal Facebook account, (b) they usually share content from their Facebook account, (c) they use this social network to communicate with their family and friends, (d) they have a smartphone, (e) they have 200 or more friends on the social network, and (f) no longer professionally active. Data saturation was got with 9 participants.

**Data collection and analysis**. Data was collected through in-depth interviews. This instrument is used in qualitative works, particularly those that seek to know the subjective experiences, feelings or beliefs of people (Crouch and McKenzie, 2006), and to know the phenomenon as it is lived by the participants (Englander, 2012). Participants were contacted through phone calls and emails. An enquiry guide with semi-structured questions was developed for each interview (Table 1).

The interviews were conducted by two researchers of the study, through conversations via digital media (phone call or video chat application). The interviews were digitally recorded and transcribed by the researchers. The average length of each interview was 40 minutes. An initial in vivo coding technique was followed to identify the participants' perspective in their own voice (Saldana, 2013). The codes were then analysed through a process of thematic analysis. Thematic analysis allows for the identification and reporting of the phenomenon studied, by recognising patterns and consistencies in the data, in which recognized themes become the categories of analysis (Braun & Clarke, 2012; Buetow, 2010).

Table 1. Interview questioning guide

| **Topics** | **Questions** |
|---|---|
| Sharing content on Facebook | For what purpose do you share a news item or post on Facebook; what type of content do you share (private or public mode); do you verify the authenticity of what you share? |
| Identifying fake news | Do you consider that the information published in a media outlet you recognise is a guarantee of the veracity of a news item; do you trust social networks as a means of information just as you trust traditional media; do you think that it is easier for young people to recognise false news on social networks? |
| Fake news on Facebook | Do you find it difficult to recognise a fake news story on Facebook; what features of the content help you recognise a fake news story on Facebook; what features could Facebook have to assist in better recognising a fake news story? |

Concerning the ethical aspects of the study, anonymity and confidentiality of participants' information was maintained by assigning them a letter code. All participants signed an informed consent form prior to the interview, and all queries related to the study were resolved. No participant declined to continue with the study or requested to withdraw the information provided. No incentives or compensation were offered to interviewees for their participation.

**Results**

The analysis allowed the identification of four themes, which summarise the different ways in which participants experience and relate to the social network and fake news.

**Theme 1: "What I have learned with my family".**

The category describes the influence that the participants' close family environment has on their decision to have a Facebook account, besides the relevance of the opinion of the surrounding members regarding their behaviour on the social network. Learning occurs through the intervention of a family member or close person, as well as self-learning through iterations or trial and error.

Participants expressed that their sons and daughters motivated them to get an account on the social network. Sometimes, it was so that they could reconnect with family members or old friends with whom they were no longer close. One interviewee expresses the reason it was recommended to her:

> *"from the time I separated [from partner], I was very lonely (...) and as my son grew up, he saw me alone and wanted me to integrate with friends at school, at university, because I became completely isolated, that's why my friends found me".*

The children also guided them on how to deal with Facebook, including guidance on how to avoid posting a news item that might be false. They also encouraged the development of a critical mindset, to make them question their future postings. One interviewee commented: *"I published that [news] and they told me 'what do you publish [that] for'?".* Among the baby bombers there is a learned lesson from sharing misleading information, a desire not to make the mistake again, because the positive opinion or approval of their friends on Facebook is important to them.

As for the friend requests the interviewees accept, they say that they are from family members, friends from school, ex-colleagues and/or neighbours. One interviewee thanked Facebook for recovering friendships: *"I thank it, I swear, I am very sincere with social networks because many people I had not seen for many years and who have been important to me, I can talk to them again".* Another participant talks about the requests she receives on Facebook: *"I could only accept if it is a schoolmate with whom I have had some contact and nothing more".* Concerning sending friendship requests, sometimes they do not do so unless they very much want to contact a friend from their youth.

**Theme 2: "Sometimes I can't recognise it".**
This category frames the fact that participants are aware of the responsibility to filter and not distribute misleading information, and the limitations they perceive to have in identifying fake news.

One interviewee explains her way of proceeding: *"Sometimes, when I don't understand, I don't share. I don't share things like that, that I don't know".* Another participant recounts her experience when information gives rise to certain doubts as to its veracity: *"It's not that I don't trust it, but I read and there are some things that leave me in doubt. But I don't find out if it's the truth, because that would already get into inquiring".*

Sometimes participants become aware that a news item is false when someone around them points it out (a reality contained in Theme 1), evidencing the limits of their ability to identify false information and the awareness they are developing based on experience:

> *"There are times when I get it wrong and my nieces tell me: 'No, auntie, that's false'. Well, that's how it happens to me sometimes, plus I may think it's OK for me, but other people point out that it's not".*

Making mistakes and then learning from the experience enhances their skills in dealing with informative content on social media. Participants also experience that sometimes their children have little patience to guide them in the learning process.

> *"On Facebook I don't share news that I can't give as real, with authenticity, they send me a lot and tell me to share, and I have to read. Generally, I don't manage to share the news they give me because I don't know if it's real".*

**Theme 3: "If I see that it is real, I can share".**
Participants state that they have critical filters that they apply to the information they find in the news media. One interviewee claim:

> *"The news is sometimes biased with situations that exist, and sometimes you say whether it is or is not real. When I analyse everything, I see if it is real or not, or if they are trying to put in things that are not, so that the people or I can trust what they say".*

The interviewees put a news item on trial when they are not sure of its veracity or objectivity, and on some occasions, they try to verify it on their own, cross-checking information from other media or using Google to get more data: *"I look on Google, on some information site to see if it is true or not".* Among the participants, identifying and trusting the source is an important element in determining whether or not information is reliable. One of them states: *"If the news is verified, I look at the source, I always look at the source, because sometimes people share anything".* Another says: *"I would have to look at the origin of that news, if [the origin of the information] is not reliable. Look for the source."*

Familiarity with social media makes it easier for interviewees to recognise if a news item is fake. One interviewee shares her experience:

> *"Sure, I have read, so it seems to me that it can work, because it has logic. I have reviewed it and I have implemented it, and it really works, and it is favourable. So, I'm happy to share things that can be useful to others on my Facebook".*

How Facebook allows content to be published gives the interviewees a perception of authenticity in the information they find, as it gives them the space to say what they believe the mainstream media does not say. It also gives them freedom in terms of the content they can share. One participant shares her experience: *"[I trust] a little more [social networks], because the traditional media are already in favour of a government"*.

Being able to debate or simply express their opinion by commenting on a news item on Facebook is valued because it gives them a space to expose and argue their points of view. One of them states: *"when something comes up on Facebook, a news item on any topic (...) and I want to have an opinion—people can have an opinion, and I sometimes have an opinion"*. They also certify that Facebook does not have filters, which is why some people *"post so much rubbish"*, although they argue that "you can't prohibit [people] from expressing what they feel".

**Theme 4: "They don't analyse"**
Participants perceive that some young family members and acquaintances do not always have the right framework to identify fake news or biased information. They describe that the lack of critical judgement among young individuals is not conducive to processing information objectively.

Participants make this observation regarding a weak foundation that is not conducive to recognising and managing information in the face of fake news. One interviewee's statement sums up the issue: *"they believe everything. It seems to me that some of them don't analyse [the information they receive], don't give it importance and post it"*. Also, regarding the absence of critical judgement, one interviewee argues that:

> *"Young people generally read everything that is put to them, sometimes they answer, sometimes they don't answer, according to what I see, but they don't analyse and sometimes they say things and well, everyone knows the behaviour of each young person"*.

Another participant perceives how some young people do not show interest in informing themselves about current affairs, which they consider being a negative characteristic. In contrast - also negative - they pay attention to news about show business personalities:

> *"I see that some young people watch the news about what happens in the country or internationally, but there are also young people who don't, they are not very aware, sometimes you talk to them and they say they don't know or haven't seen it. But they watch showbiz (...) on TV, that they are aware of"*.

**Discussion**
This study analyses the media literacy process that baby boomer women experience when confronted with fake news on Facebook, and how they perceive trustworthiness regarding a source, and identifies misleading information on the social network. Previous research argues that women rarely analyse news critically (Adolfsson, et al, 2017). The findings of this paper do not agree with this statement. The findings broaden the scarce knowledge about how baby boomer women deal with fake news, as well as their perceived responsibility not to promote its dissemination. The findings highlight their critical thinking skills, empirically developed based on the knowledge they gain from their family environment. These results are consistent with those of Flintham et al (2018), who report that simply being aware of the possibility that a news story may not be authentic increases new media literacy. This study contributes to the academic discussion regarding digital media literacy, and the development of critical skills in the face of misleading information among baby boomer populations, for whom this work can lead to greater well-being in their lives.

The theme 'What I have learned with my family' shows that family members are a motivator for opening a Facebook user account. This is in line with the observations of Randall, Pauley & Culley (2015) that women report their children have a powerful influence on their decision to use the social network. The same category makes it visible that the participants receive help and guidance on how to use Facebook from younger family members in their close environment. These family members instruct them on how to identify fake news on the social network. This phenomenon is a reverse socialisation process, as it describes a situation where older people learn from and are socialised by young people (McClain, 2011). Intimate family members are made up as agents of motivation and literacy, transferring their knowledge about Facebook use and misleading information. These agents help them internalise their own and other family members' experiences of posting misleading information on Facebook. The transfer of experiences fosters the development of critical thinking, which then allows participants to question whether it is appropriate (or not) to post a news item on Facebook, or whether it is a requirement to seek sources to confirm the information. Identifying sources is also part of the learning process gained.

The interviewees reported using Google or consulting other people close to them to check the data to determine its reliability. This learning is relevant because, as Scolari (2016) suggests, literacy cannot be limited to physical media. This literacy required by women over 65, who have been confronted with various technological changes and exposure to a large number of media (Guess, et al, 2019), will allow them to have a better quality of life (Scolari, 2016), preventing them from having a bad time and being negatively singled out by their friends on Facebook for sharing false information.

Learning is a constant endeavour, according to the interviewees, and they turn to their close family members when they are unsure or want to confirm a piece of news. They are also seeking to learn how to use their social networks. In addition, they are always willing to learn and receive feedback from their close environment, to go to the source of the news or publication and verify if what they are going to share is accurate. "Sometimes I can't recognise it' is a theme that reflects the awareness among participants that fake news is not only presented to them, but that their critical ability has limitations. The theme frames the reality that participants are aware of the responsibility to filter and not distribute misleading information, and the limitations they perceive themselves to have in identifying fake news - in the words of one interviewee, "I read and there are some things that I'm kind of in doubt about". The theme "If I see it is real, I can share" refers to the same experience, in this case when the person experiences that their critical ability is adequate to identify the reliability of information, and acts accordingly. This is an experience in which participants express awareness and reliability in their capability to identify the trustworthiness of information. They believe that the identification of a source they recognise, and the perception of the internal and external logic of an information, are elements that make it easier for them to practice critical judgement.

Throughout this study, researchers observe that Baby Boomer women have been accompanied since the creation of their Facebook account by their offspring and close relatives, to guide them in using this social network, identify fake news and learn how to improve their skills in detecting reliable sources. In addition, they have learned from previous experiences, originating from sharing fake news, as it has affected them emotionally. This is because they expect to always have approval of their behaviour on Facebook from their friends. They are also distrustful of the traditional national media and value above all the freedom of expression that social networks allow their users, even though this may allow some people to misinform, even intentionally, through their opinions or by sharing misleading information.

The findings of this research are limited to the self-reported experiences of the interviewees, which occur from their individual frames of reference. These findings cannot rule out the existence of other elements that may have influenced the way they conduct themselves when processing information. For example, receiver characteristics as subjects their attitude towards the media, such as trustworthiness with which they deal with diverse events in their lives, or development of their analytical skills. Even their level of engagement with the events covered by the media in the local and international news broadcasts. Further research can examine the impact of these variables on behaviour. Including the mediating effect that the availability of time or interest of children, or other close relatives, may have in assisting members of this generation in dealing with misleading information on social media. This research was conducted among women who had a negative perception of sharing fake news and were not motivated to spread such stories. Further studies could investigate the identification skills of baby boomers who consciously distribute fake news for a variety of motivations.